# Doublet-triplet energy transfer dominated photon upconversion

Jianlei Han, [†] Yuqian Jiang, [†] Ablikim Obolda, [†] Feng Li*, Minghua Liu* and Pengfei Duan*

Mr. J. Han, Dr. Y. Jiang, Prof. Dr. P. Duan, Prof. Dr. M. Liu

CAS Key Laboratory of Nanosystem and Hierarchical Fabrication

National Center for Nanoscience and Technology (NCNST)

No.11 ZhongGuanCun BeiYiTiao, 100190 Beijing, P.R. China

E-mail: duanpf@nanoctr.cn, liumh@iccas.ac.cn

Mr. A. Obolda, Prof. Dr. F. Li

State Key Laboratory of Supramolecular Structure and Materials，College of Chemistry

Jilin University

Qianjin Avenue 2699, Changchun, 130012, P. R. China

E-mail: lifeng01@jlu.edu.cn

Prof. Dr. M. Liu

Beijing National Laboratory for Molecular Science (BNLMS), CAS Key Laboratory of Colloid, Interface and Chemical Thermodynamics

Institute of Chemistry, Chinese Academy of Sciences

No. 2 Zhongguancun BeiYiJie,100190 Beijing, P.R. China

Mr. J. Han

School of Chemical Engineering and Technology

Tianjin University

Tianjin, 300072, P. R. China

[†] these authors contributed equally to this work

**Abstract:** Stable luminescent π-radicals with doublet emission have aroused a growing interest for functional molecular materials. We have demonstrated a neutral π-radical dye (4-N-carbazolyl-2,6-dichlorophenyl)bis(2,4,6-trichlorophenyl)-methyl (TTM-1Cz) with remarkable doublet emission, which could be used as triplet sensitizer to initiate the photophysical process of triplet-triplet annihilation photon upconversion (TTA-UC). Dexter-like excited doublet-triplet energy transfer (DTET) was confirmed by theoretical calculation. A mixed solution of TTM-1Cz and aromatic emitters could ambidextrous upconvert red light ($\lambda$ = 635 nm) to cyan light or green light ($\lambda$ = 532 nm) to blue light. This finding of DTET phenomena provides a new perspective in designing new triplet sensitizer for TTA-UC.

Organic radicals have abundant applications such as spintronics,[1] polarizing agents,[2] organic magnetism[3] and accelerating chemical reactions[4] because their unpaired electron can easily take part in physical processes and chemical reactions. Generally, neutral radicals are quite unstable. However, through the molecular design, stable neutral radicals can be obtained and they are able to withstand oxygen and light for an extremely long time at room temperatures.[5] Recently luminescent neural π-radicals[6] have aroused a growing interest due to their special emission from doublet exciton, which is different from the emission of closed-shell molecules (from either singlet or triplet exciton, Figure S1). Thus they can be used to realize the particular functions which cannot be achieved by the closed-shell luminescent molecules. One example is that a luminescent neural π-radical was successfully used as the emitter of an organic light-emitting diode (OLED). Because the emission of the OLED comes from the double exciton of the π-radical whose transition to the ground state is spin-allowed, the transition problem of triplet exciton of closed-shell emitters is thus circumvented [7].

Photo upconversion (UC) through annihilation between two-long-lifetime excited triplets (TTA) has recently come into the spotlight because of its wide applications that range from renewable energy productions to bioimaging and phototherapy. It attracted much attention due to its occurrence with low-intensity and non-coherent incident light.[8] Figure S2 shows a typical scheme for TTA-based UC, in which a molecular sensitizer (triplet donor)-emitter (triplet acceptor) pair shares roles. A triplet excited state ($T_1$) of the donor is formed via intersystem crossing (ISC) from the singlet excited state ($S_1$) of the donor. Donor-to-acceptor (D-A) triplet-triplet energy transfer (TTET) populates the $T_1$ state of acceptor, and TTA between two acceptor triples produces a higher energy acceptor $S_1$ state that consequently emits upconverted delayed fluorescence. Generally, metal organic complexes were commonly used as the triplet donor due to the high triplet quantum yield ($\Phi_{ISC} \approx 100\%$). However, the energy loss $\Delta E_{ST}$ during ISC, typically hundreds of meV, is contradict to the efficient low energy photon upconversion (Figure S2).[9] On the other hand, developing pure organic sensitizer to replace the metal organic complex is also an important issue in TTA-based photon upconversion research field.[10]

Here, we report a solution for overcoming the energy loss issue during ISC and replacing the metal organic complex sensitizer with pure organic π-radical dye in TTA-UC. This is based on excited doublet-triplet energy transfer (DTET) by directly exciting an organic π-radical, (4-N-carbazolyl-2,6-dichlorophenyl)bis(2,4,6-trichlorophenyl)methyl (TTM-1Cz).[6a, 6b] We provide definitive experimental evidences that doublet energy transfer proceeds rapidly and efficiently from excited radical dye TTM-1Cz to the acceptor. Dexter-like doublet-triplet energy transfer could be observed in TTA photon upconversion system.

Considering the essential differences with fluorescence (singlet exciton emission) and phosphorescence (triplet exciton emission), we term the doublet exciton emission as "radicorescence" which is the combination of "radical" and "luminescence". We hope this expression could represent the luminescence properties of this class of compounds. In this work, we tried to use radical dye as an energy donor to initiate a TTA-based photon upconversion photophysical process. As shown in Figure 1a, through a direct excitation of a radical dye, doublet exciton ($^2D^*$) could sensitize an acceptor accompanying with doublet-triplet energy transfer. Different from the typical triplet exciton sensitized TTA-UC, the excited doublet exciton could directly sensitize the acceptor without energy loss resulted from intersystem crossing. This approach will open up a new research field about triplet sensitizer designing.

It is well known that triplet-triplet energy transfer follows the electron exchange mechanism (Dexter) with the characteristic short triplet interaction distance (*ca.* 1nm).[11] Here, electron-exchange-based mechanism also could be afforded to doublet-triplet energy transfer (Figure 1b):

$$^2D^* + {}^1A \rightarrow {}^2D + {}^3A^* \qquad (1)$$

The energy transfer integral of such energy transfer process can be expressed as:

$$V_{DA} = \langle \psi_{^2D^*}\psi_{^1A} | \hat{H} | \psi_{^2D} \psi_{^3A^*} \rangle$$
$$= \langle \psi_r | \hat{H} | \psi_p \rangle \qquad (2)$$

$\Psi_r$ and $\Psi_p$ are the spin-localized wave functions before and after energy transfer, respectively. To simplify the problem, we assume that $\Psi_r$ and $\Psi_p$ are composed of the same set of core orbitals, and the differences are only in the four highest occupied spin-orbitals, as shown in Figure 1b. Thus, we can express $\Psi_r$ and $\Psi_p$ in Slater determinant:

$$\psi_r = |\psi_{core}\phi^{\alpha}_{D,SUMO}\phi^{\alpha}_{A,HOMO}\phi^{\beta}_{A,HOMO}| \qquad (3)$$

$$\psi_p = |\psi_{core}\phi^{\beta}_{D,SOMO}\phi^{\alpha}_{A,HOMO}\phi^{\alpha}_{A,LUMO}| \qquad (4)$$

So, the coupling can be derived as

$$V_{DA} = \left[\phi^{\alpha}_{D,SUMO}\phi^{\beta}_{A,HOMO} \| \phi^{\beta}_{D,SOMO}\phi^{\alpha}_{A,LUMO}\right]$$
$$= \left[\phi^{\alpha}_{D,SUMO}\phi^{\beta}_{A,HOMO} | \phi^{\beta}_{D,SOMO}\phi^{\alpha}_{A,LUMO}\right] - \left[\phi^{\alpha}_{D,SUMO}\phi^{\alpha}_{A,LUMO} | \phi^{\beta}_{D,SOMO}\phi^{\beta}_{A,HOMO}\right]$$
$$= -\left[\phi^{\alpha}_{D,SUMO}\phi^{\alpha}_{A,LUMO} | \phi^{\beta}_{D,SOMO}\phi^{\beta}_{A,HOMO}\right] \qquad (5)$$

where $\phi^{\alpha(\beta)}$ is the occupied spin-orbital for donor or acceptor as shown in Figure 1b. It shows the same expression as triplet-triplet energy transfer,[12] which is completely the exchange electronic integral. Therefore, when a stable radical dye is used as an energy donor, the corresponding excited doublet exciton can transfer energy to an acceptor through electron exchange by producing an excited triplet exciton and a ground state of donor. In this work, we have used a neutral π-radical dye TTM-1Cz as a triplet sensitizer to initiate TTA-UC by sensitizing two different acceptors bis(phenylethynyl)anthracene (BPEA) and 9,10-diphenylanthracene (DPA) (Figure 2a). BPEA and DPA have been widely reported in TTA-UC systems as two kinds of typical triplet annihilators sensitized by metal complexes. Here, we firstly tried to sensitize these two annihilators by using a radical dye TTM-1Cz. It has been reported that TTM-1Cz exhibited strong broad doublet exciton emission in the near-infrared (NIR) region (600-800 nm) which could be used as brand-new emitter in organic light-emitting devices.[7] A toluene solution of TTM-1Cz exhibited a broad absorption band centered at approximately 600 nm which is assigned to the electronic transition from SOMO to SUMO (Figure 2b). The PL spectrum (centered at 680 nm) demonstrated that the emission was originated from the transition of SUMO to SOMO, that is, the radiative decay from the doublet excitons. The acceptor BPEA shows intense π-π* absorption bands at 438 and 458 nm with corresponding fluorescence ranging from 460 to 620 nm. The absorption of TTM-1Cz reveals a small overlap with the emission of BPEA which is negligible for UC. It should be noted that, due to the broad absoprtion (500-650 nm) of TTM-1Cz, DPA is also a candidate for doublet annihilator. Compared the absorption and emission between DPA and TTM-1Cz, it is expected these dye pairs have the potential to achieve UC (Figure S3a). In this work, we will carefully demonstrate the UC properties of TTM-1Cz/BPEA pairs while TTM-1Cz/DPA pairs will be posted in the supporting informations.

As shown in Figure 3a, by sensitizing the BPEA triplet with TTM-1Cz, the incident red light (λ = 635 nm) is successfully upconverted to the cyan light (λ = 507 nm) in solution (Figure 3a). Steady-state luminescence spectra at varied incident laser power clearly showed the upconverted emission of BPEA at 507 nm. Moreover, due to the broad absorption band from 500-650 nm of TTM-1Cz, it also could sensitize another acceptor DPA triplet with the incident green light (λ = 532 nm) and the upconverted blue light (λ = 440 nm, Figure S3b). The UC emission intensity from the TTM-1Cz/BPEA mixed solution was plotted as a function of excitation intensity (Figure 3b). A slope change from 2 to 1 was clearly observed in double-logarithmic plots of the UC emission intensity of solution against excitation intensity.[13] Generally, the excitation intensity dependence of UC emission intensity changes from quadratic to linear above a threshold excitation intensity ($I_{th}$). Above $I_{th}$, TTA becomes the main deactivation channel for the acceptor triplet, and consequently the UC quantum yield shows saturation.[13] The cyan UC emission showed excellent photophysical stability, as confirmed by the good maintenance of UC emission intensity after continuous excitation over 3000 s (laser intensity = 210 mW cm$^{-2}$) (Figure S4a). Even under very high laser intensity (1810 mW cm$^{-2}$), the UC emission showed slight decrease then reached to the stable state. We have carefully retested the electron paramagnetic resonance (EPR) spectrum of TTM-1Cz/BPEA mixture after long time laser irradiation (Figure S4c). The unpaired electron could be clearly observed which indicated that the radical sensitizer possesses remarkable stability under strong laser irradiation. The UC pairs TTM-1Cz/DPA also exhibited excellent stability in UC process (Figure S4b and 4d). We also have measured the emission of pure TTM-1Cz in deaerated toluene under laser irradiation, in order to completely confirm the authenticity of doublet sensitized TTA-UC (Figure S5). By applying strong laser irradiation (635 nm) in pure TTM-1Cz deaerated toluene solution, no UC emission could be observed while only doublet exciton emission of TTM-1Cz around 680 nm could be detected. Similar phenomenon could be obtained by shining the green laser 532 nm. These results further confirmed that TTM-1Cz sensitizer TTA-UC should be definitely occurred.

In order to evaluate the efficiency of DTET, we compared the radicoresecence lifetime of TTM-1Cz with ($\tau_{D'}$) and without ($\tau_D$) the acceptor BPEA (Figure 3c). The energy transfer efficiency $\Phi_{DTET}$ could be calculated with the equation $\Phi_{DTET} = (\tau_D - \tau_{D'})/\tau_D \times \%$ to give a value 46%. This indicates that the energy transfer from doublet exciton to acceptor triplet is not that efficient which might be due to the short lifetime of doublet exciton of TTM-1Cz ($\tau_{TTM-1Cz}$ = 27 ns). We then quantified the TTA-UC quantum yield ($\Phi_{UC}$) of the TTM-1Cz/BPEA pairs in deaerated toluene by using methylene blue (MB) as a reference defined as the ratio of emitted photon numbers to absorbed photon numbers, and thus the theoretical maximum of the TTA-UC quantum yield ($\Phi_{UC}$) is 50%.[14] Meanwhile, in many reports this value is multiplied by 2 to set the maximum conversion efficiency at 100%. To avoid the confusion between these different definitions, the UC quantum yield is written as $\Phi'_{UC}$ (= $2\Phi_{UC}$) when the maximum efficiency is standardized to be 100%. With increasing the excitation intensity, the $\Phi'_{UC}$ value increased to 0.02% then it reached saturation (Figure S6). The low UC quantum yield, originating from the low TTM-1Cz/BPEA DTET efficiency, should be due to the short doublet lifetime of TTM-1Cz.

The TTA-base UC mechanism was confirmed by the lifetime and excitation intensity dependence of the upconverted emission. The UC emission at 507 nm showed a microsecond-scale decay profile, which is ascribed to the mechanism based on long-lived triplet species (Figure 3d). Such a long decay was absent in single-component solution of TTM-1Cz or BPEA. A triplet lifetime of acceptor BPEA ($\tau_{A,T}$) was estimated as 138 $\mu$s by tail fitting based on the knows relationship $I_{UC}(t) \propto \exp(-2t/\tau_{A,T})$.[15] In this time domain, the annihilation efficiency becomes negligible compared with spontaneous decay of the triplet and thus the $\tau_{A,T}$ value can be simply estimated. The remarkable long triplet lifetime of acceptor enabled the efficient photophysical process of triplet-triplet energy transfer and triplet-triplet annihilation.

For illustrating the rationality of experiments, we apply quantum mechanics and molecular dynamics to study the doublet-triplet energy transfer process in TTM-1Cz/BPEA system as well as TTM-1Cz/DPA (SI part III). The Franck-Condon integrals for $D_1 \rightarrow D_0$ emission for TTM-1Cz and $S_0 \rightarrow T_1$ absorption for BPEA (Fig. S8) exhibited partial overlap which ensures the occurrence of energy transfer. Considering the electron exchange is short distance interaction, doublet-triplet energy transfer can only happen when donor and acceptor are nearby. Here, we simply selected one snapshot with the shortest centroid distance as example to study the energy transfer property, since the calculation of statistic averages during the equilibrium process which can represent the experimental observed values is too expensive and unnecessary. By using direct coupling method, the energy transfer integral for TTM-1Cz/BPEA dimer with the shortest centroid distance was achieved as only 2.0 meV, which accords with the magnitude of electronic exchange integral obtained in triplet-triplet energy transfer.[12] Then, with Fermi's Golden Rule, the doublet-triplet energy transfer rate in such TTM-1Cz/BPEA dimer is $5.82 \times 10^{10}$ s$^{-1}$ which is larger than the observed value ($3.18 \times 10^7$ s$^{-1}$). Since the electronic exchange integral is exponentially decrease with the increase of distance, the relatively larger energy transfer rate in the dimer with the shorted distance ensure the happening of DTET between TTM-1Cz and BPEA. The upconversion pair TTM-1Cz/DPA also exhibited the same behavior under the same calculation (SI part III).

To scrutinize the grounds of the $I_{th}$ value, we determined all the related parameters shown in the expression: $I_{th} = 1/\alpha \Phi_{ET}(\tau_{A,T})^2 \gamma_{TT}$.[13, 16] Where $\alpha$ is the absorption coefficient at an excitation wavelength of 532 nm, $\Phi_{ET}$ is the donor-to-acceptor DTET efficiency, $\tau_D$ is the lifetime of an acceptor excited triplet, and $\gamma_{TT}$ is the second-order annihilation constant for TTA. For the selected concentration of [TTM-1Cz] = 0.1 mM and [BPEA] = 20 mM, $\alpha$ was as high as 0.31 cm$^{-1}$. The $\Phi_{ET}$ value was calculated as 46%. The acceptor triplet lifetime was obtained by the tail-fitting of the UC emission decay, and it was reasonably long (138 μs). The annihilation constant $\gamma_{TT}$ can be calculated as $1.3 \times 10^{-10}$ cm$^3$ s$^{-1}$. On the other hand, $\gamma_{TT}$ can also be expressed as $\gamma_{TT} = 8\pi D a_0$, where $D$ is the triplet diffusion coefficient and $a_0$ is the effective triplet-triplet interaction distance. By suing the Stokes-Einstein relationship $D = 3k_BT/6\pi\eta R_A$, the viscosity of the solvent ($\eta$, 0.58 cP), and the molecular radius of the acceptor BPEA ($R_A$, 3.1 Å), the $D$ value was calculated to be $3.8 \times 10^{-5}$ cm$^2$ s$^{-1}$. The interaction distance of the acceptor triplet $a_0$ was determined to be 1.3 nm, which agrees with the usual triplet interaction distance of *ca.* 1 nm for the electron exchange (Dexter) mechanism. It is therefore strongly supported that all the parameters including $I_{th}$ were obtained in good accuracy.

In conclusion, we have demonstrated the first example of excited doublet exciton sensitized TTA-UC which enables upconverting red to cyan light as well as green to blue light. π-radical dye TTM-1Cz worked as triplet sensitizer will provides a new perspective in TTA-UC because it not only offers a useful methodology for reducing the energy loss during triplet sensitization but also a new kind of pure organic triplet sensitizer for avoiding tedious synthesis of expensive metal complexes. In addition, it reveals a new application for excited doublet exciton emission. This work underpins the important concept of doublet emission and stimulates the exploration of doublet-triplet energy transfer toward highly efficient TTA-UC upconverters, which would find a number of applications in many disciplines.

**Experimental Section**

Spectral measurements: UV-vis spectra were recorded on Hitachi U-3900 spectrophotometer. Fluorescence spectra were measured were obtained using and F-4500 fluorescence spectrophotometer. Upconverted emission spectra were recorded on a Zolix Omin-λ500i monochromator with photomultiplier tube PMTH-R 928 using an external excitation source, 635 nm or 532 nm semiconductor laser. The absolute fluorescence quantum yield was measured by using an absolute PL quantum yield spectrometer (Edinburg FLS-980 fluorescence spectrometer) with a calibrated integrating sphere and fluorescence lifetime measurements were recorded on the same spectrometer using time-correlated single photon counting (TCSPC), phosphorescence decays and upconverted emission decays were recorded on Edinburg FLS-980 using time-resolved single photon counting-multi-channel scaling (MCS). The quality of the fit has been judged by the fitting parameters such as $\chi^2$ (<1.2) as well as the visual inspection of the residuals. EPR measurements: EPR spectrum of the deaerated mixing solution of TTM-1Cz/acceptor was measured using the JES-FA200 EPR spectrometer at ambient temperature.

**Supporting Information**

Supporting Information is attached in this file.

**Acknowledgements**

This work was supported by National Key Basic Research and Development Program of China (Grant No. 2016YFB0401001, 2016YFA0203400) founded by MOST. The National Natural Science Foundation of China (Nos 51673050, 51673080, 21603043), "Strategic Priority Research Program" of the Chinese Academy of Sciences (XDB12020200), and "New Hundred-Talent Program" research fund of the Chinese Academy of Sciences

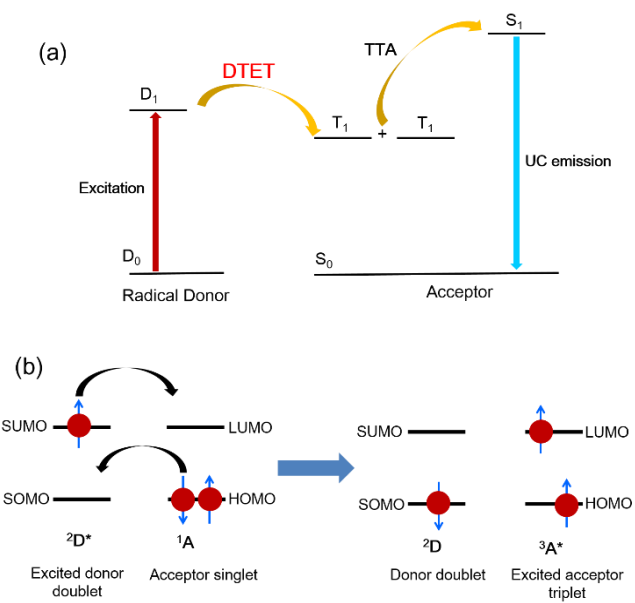

**Figure 1.** (a) Illustration of TTA-based UC sensitized by doublet exciton. (b) Mechanism of electron-exchange-based Dexter-like doublet-triplet energy transfer.

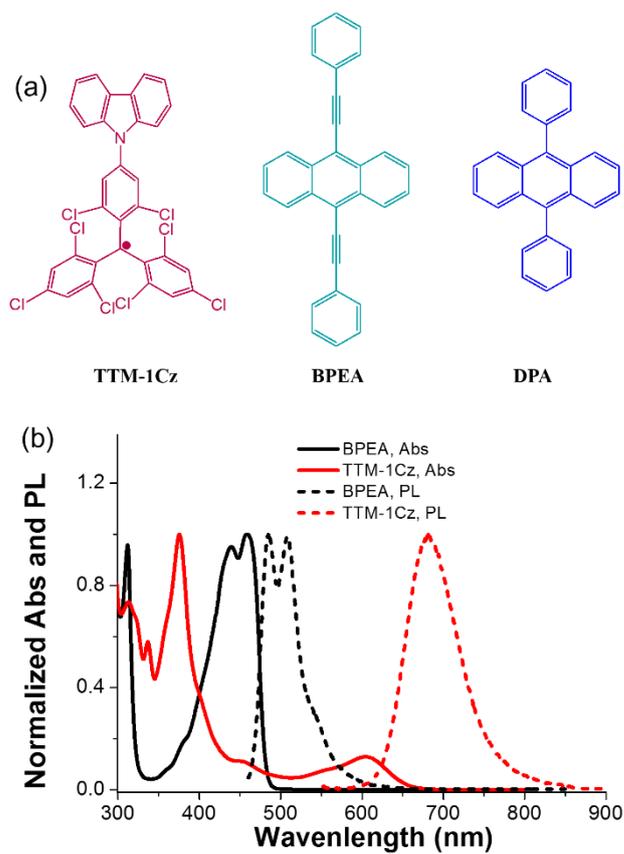

**Figure 2.** (a) Molecular structures of all the compounds used in this work. (b) Normalized absorption and emission spectra of BPEA (0.5 mM) and TTM-1Cz (0.5 mM; $\lambda_{ex}$ = 530 nm) in toluene.

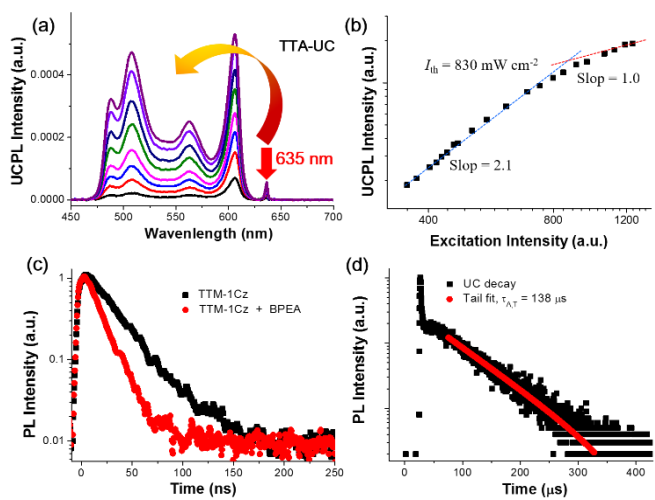

**Figure 3.** (a) Upconversion emission spectra of TTM-1Cz/BPEA with different incident power density of 635 nm laser in deaerated toluene. (b) Dependence of UC emission intensity at 507 nm on the incident power density. The dashed lines are fitting results with slopes of 2.1 (blue) and 1.0 (red) in the low and high-power regimes, respectively. (c) Time resolved doublet emission at 680 nm of TTM-1Cz with and without BPEA. (d) Time resolved upconverted emission at 507 nm of the TTM-1Cz / BPEA pair ($\lambda_{ex}$ = 635 nm). The red fitting curve was obtained by considering the relationship of $I_{UC}(t) \propto \exp(-2t/\tau_{A,T})$ where $\tau_{A,T}$ is acceptor triplet lifetime. ([TTM-1Cz] = 0.05 mM, [BPEA] = 5 mM, $\lambda_{ex}$ = 635 nm).

**The table of contents:**

The first example of neutral π-radical dye sensitized, triplet-triplet annihilation based photon upconversion is reported. Excited doublet exciton by direct exciting the radical dye could transfer the energy to a triplet acceptor through a Dexter-like doubled-triplet energy transfer (DTET). Meanwhile, the radical dye works as a versatile sensitizer to initiate the upconversion emission from red to cyan as well as green to blue.

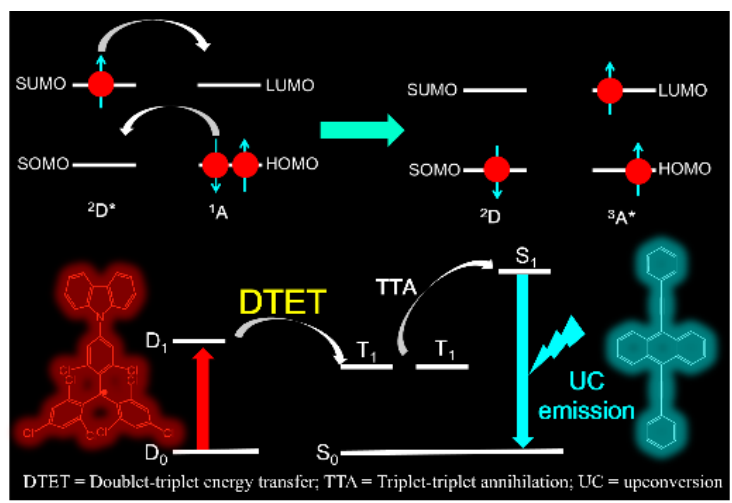

# Supporting Information

# Doublet-triplet energy transfer dominated photon upconversion

Jianlei Han, [†] Yuqian Jiang, [†] Ablikim Obolda, [†] Feng Li*, Minghua Liu* and Pengfei Duan*

Mr. J. Han, Dr. Y. Jiang, Prof. Dr. P. Duan, Prof. Dr. M. Liu

CAS Key Laboratory of Nanosystem and Hierarchical Fabrication

National Center for Nanoscience and Technology (NCNST)

No.11 ZhongGuanCun BeiYiTiao, 100190 Beijing, P.R. China

E-mail: duanpf@nanoctr.cn, liumh@iccas.ac.cn

Mr. A. Obolda, Prof. Dr. F. Li

State Key Laboratory of Supramolecular Structure and Materials，College of Chemistry

Jilin University

Qianjin Avenue 2699, Changchun, 130012, P. R. China

E-mail: lifeng01@jlu.edu.cn

Prof. Dr. M. Liu

Beijing National Laboratory for Molecular Science (BNLMS), CAS Key Laboratory of Colloid, Interface and Chemical Thermodynamics

Institute of Chemistry, Chinese Academy of Sciences

No. 2 Zhongguancun BeiYiJie,100190 Beijing, P.R. China

Mr. J. Han

School of Chemical Engineering and Technology

Tianjin University

Tianjin, 300072, P. R. China

[†] these authors contributed equally to this work

**Contents**

**Part I. Experimental details**

**Materials.**

All reagents and solvents were used as received otherwise indicated. The sensitizer (4-N-carbazolyl-2,6-dichlorophenyl)bis(2,4,6-trichlorophenyl)methyl (TTM-1Cz) was synthesized according to the reported procedures.[1] The acceptors bis(phenylethynyl)anthracene (BPEA) and 9,10-diphenylanthracene (DPA) were purchased from TCI and used as received.

For TTA-UC measurements, toluene solutions of acceptor (BPEA, DPA) and donor (TTM-1Cz) were deaerated by repeated freeze-pump-thaw cycles.

**Determination of TTA-UC quantum yield by relative method**.

The upconverted emission quantum efficiency (UC) was determined relative to a standard according to the following equation,[2]

$$\Phi_{UC} = \Phi_{std} \left(\frac{A_{std}}{A_{uc}}\right)\left(\frac{I_{uc}}{I_{std}}\right)\left(\frac{\eta_{uc}}{\eta_{std}}\right)^2$$

where, Φ, A, I and η represent a quantum yield, absorbance, integrated photoluminescence spectral profile, and refractive index of the solvents used as a standard, respectively. The subscripts UC and std denote the parameters of the upconversion and standard systems. The UC quantum yield was determined relative to a standard, methyl blue in ethanol ($\Phi_{std}$ = 3%) under 635 nm excitation.

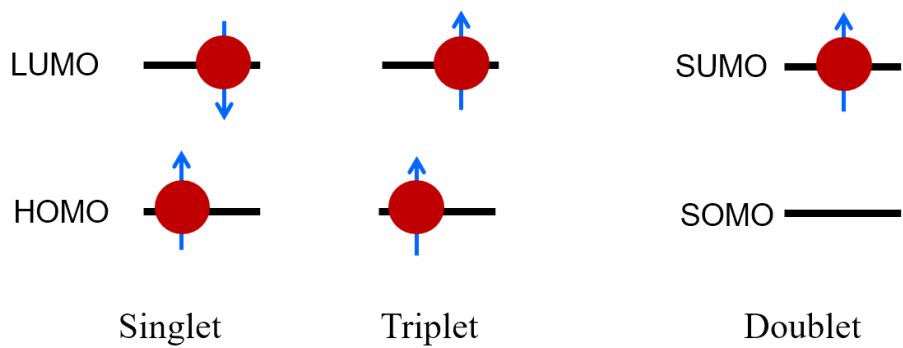

**Figure S1**. Illustration of electron spin state of different excitons.

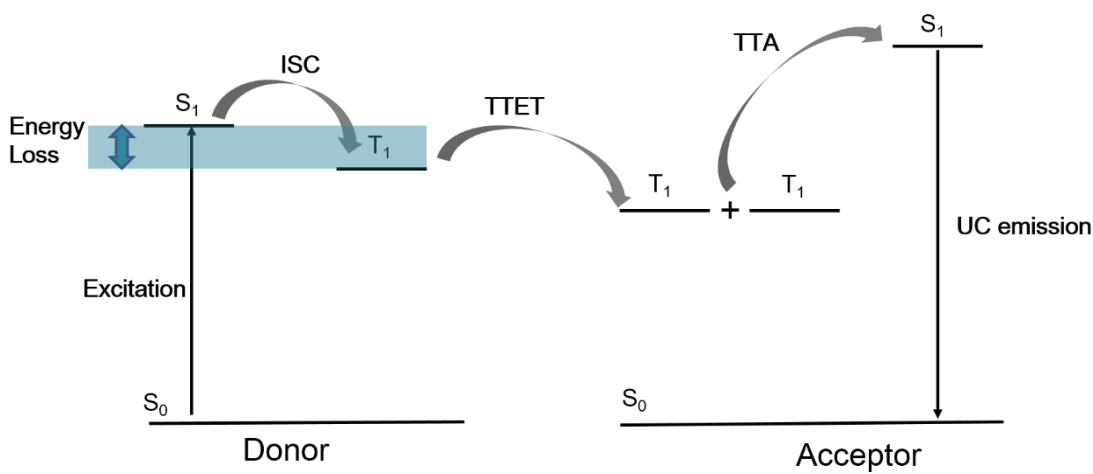

**Figure S2.** Common TTA-based UC mechanism *via* donor ISC that involves energy loss from the energy difference of $S_1$ and $T_1$ of the sensitizer (S = singlet, T = triplet). The TTA-UC involves a donor (sensitizer) with high intersystem crossing (ISC) efficiency and an acceptor (emitter) with high fluorescence quantum yield. First, the donor absorbs the low energy light to produce the excited singlet state ($S_1$). Second, the triplet state ($T_1$) of donor is populated through ISC. Third, triplet-triplet energy transfer (TTET) from donor $T_1$ to the triplet state of acceptor *via* the Dexter mechanism. Finally, the collision and annihilation (TTA) between two acceptor triplets produce a higher-energy singlet excited state of the acceptor $S_1$ which radiates upconverted delayed fluorescence. (a) Molecular structures of all the compounds used in this work.

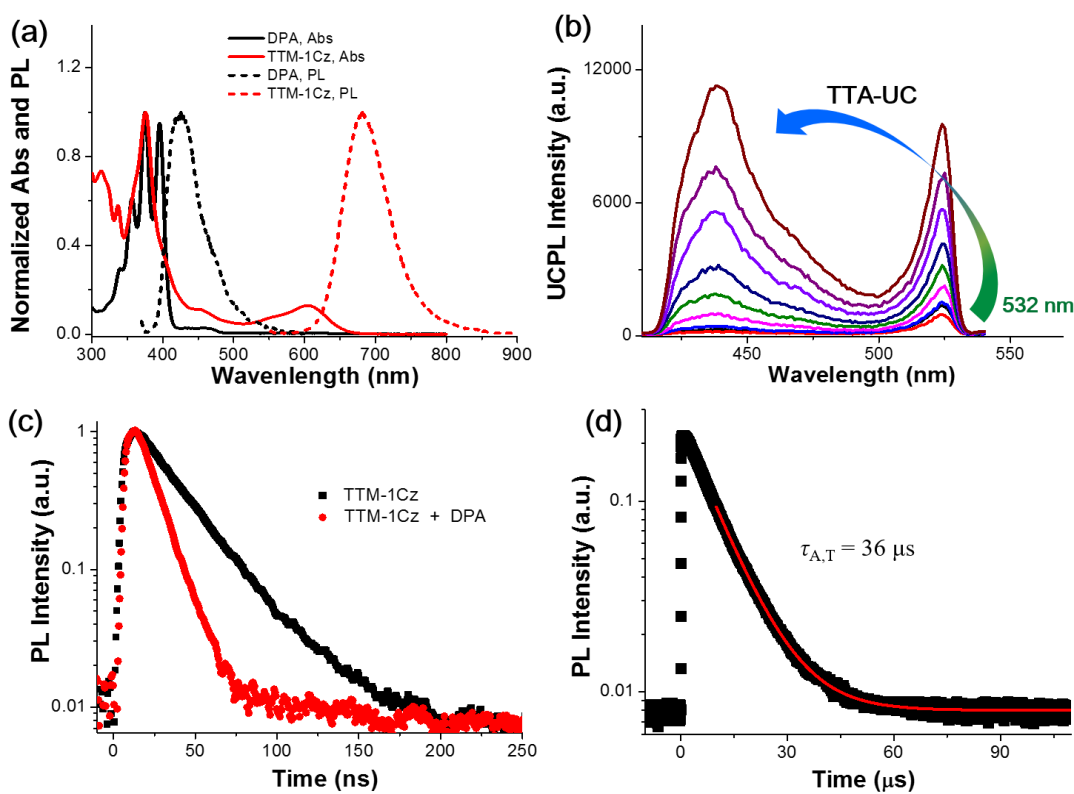

**Figure S3.** (a) Normalized absorption and emission spectra of DPA (0.1 Mm, $\lambda_{ex}$ = 375 nm) and TTM-1Cz (0.1 mM; $\lambda_{ex}$ = 532 nm) in toluene. (b) Upconversion emission spectra of TTM-1Cz/DPA with different incident power density of 532 nm laser in deaerated toluene. (c) Time resolved doublet emission at 700 nm of TTM-1Cz with and without

DPA. (d) Time resolved upconverted emission at 440 nm of the TTM-1Cz / DPA pair ($\lambda_{ex}$ = 532 nm). The red fitting curve was obtained by considering the relationship of $I_{UC}(t) \propto \exp(-2t/\tau_{A,T})$ where $\tau_{A,T}$ is acceptor triplet lifetime.[3] ([TTM-1Cz] = 0.1 mM, [DPA] = 20 mM, $\lambda_{ex}$ = 532 nm)

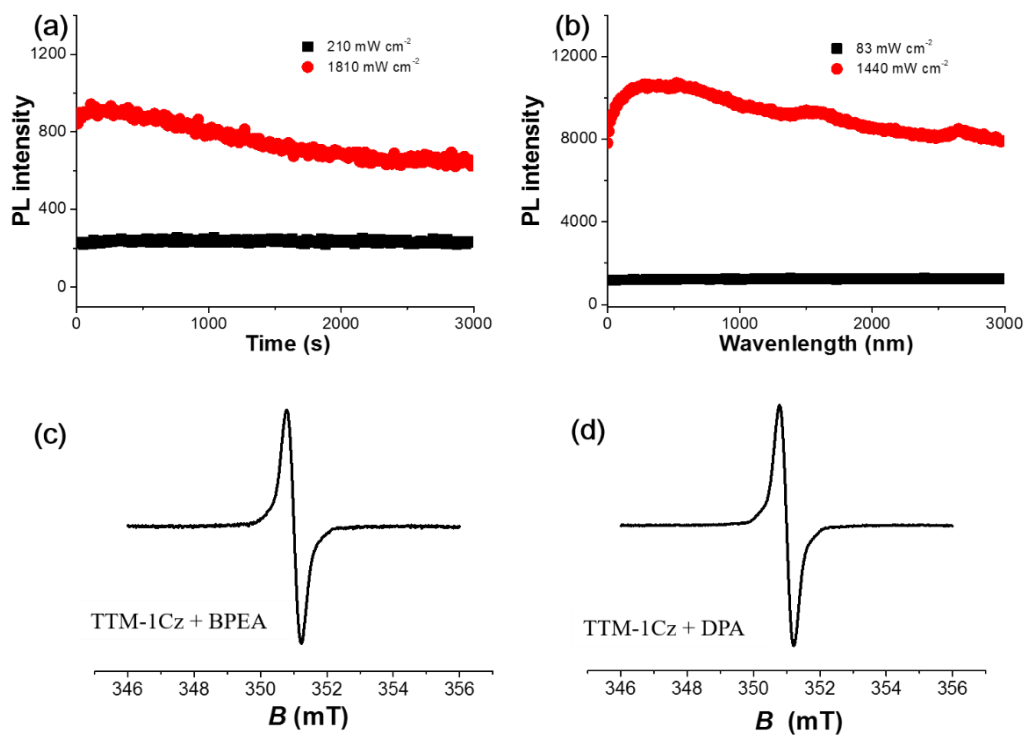

**Figure S4.** (a) Time dependence of upconverted emission intensity at 507 nm of TTM-1Cz/BPEA in toluene solution ($\lambda_{ex}$ = 635 nm). (b) Time dependence of upconverted emission intensity at 440 nm of TTM-1Cz/DPA in toluene solution ($\lambda_{ex}$ = 532 nm). (c) The EPR spectrum of TTM-1Cz/BPEA in toluene solution after TTA-UC measurement. (d) The EPR spectrum of TTM-1Cz/DPA in toluene solution after TTA-UC measurement.

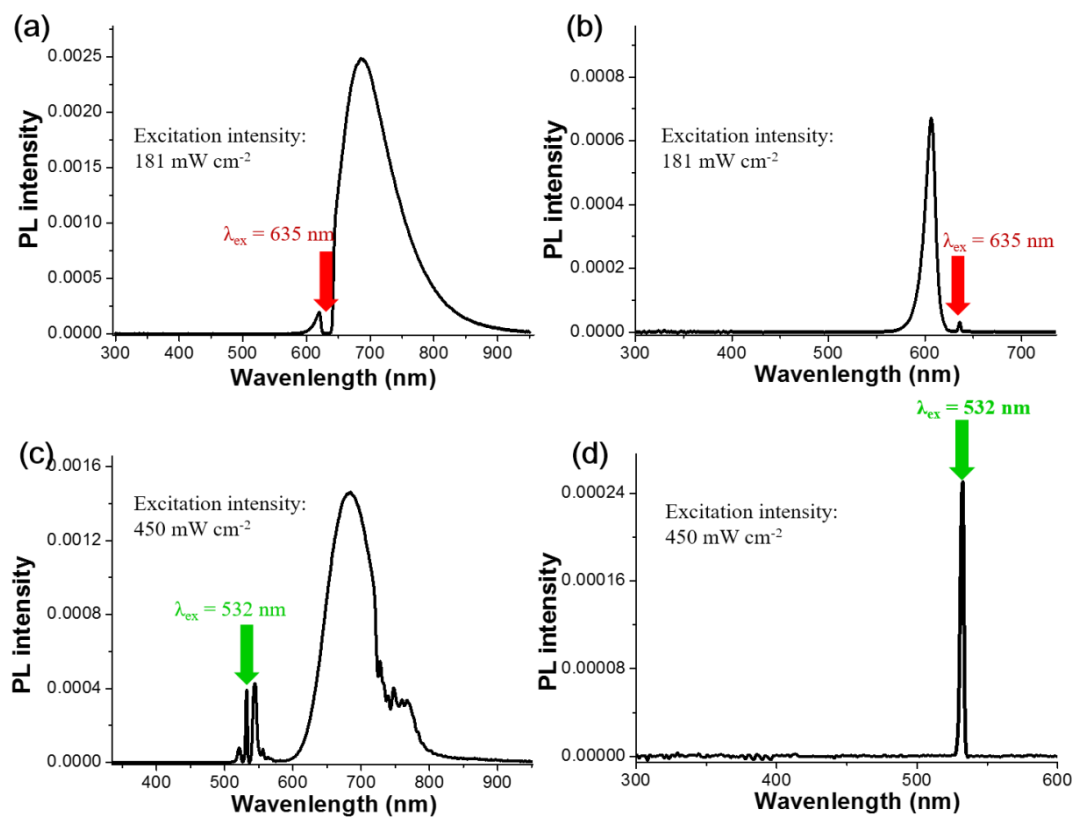

**Figure S5.** (a) Photoluminescence spectra of TTM-1Cz in toluene deaerated by repeated freeze-pump-thaw cycles ($\lambda_{ex}$ = 635 nm) with long-pass filter. (b) Photoluminescence spectra of TTM-1Cz in toluene deaerated by repeated freeze-pump-thaw cycles ($\lambda_{ex}$ = 635 nm) with short-pass filter. (c) Photoluminescence spectra of TTM-1Cz in toluene deaerated by repeated freeze-pump-thaw cycles ($\lambda_{ex}$ = 532 nm) with long-pass filter. (d) Photoluminescence spectra of TTM-1Cz in toluene deaerated by repeated freeze-pump-thaw cycles ($\lambda_{ex}$ = 532 nm) with long-pass filter.

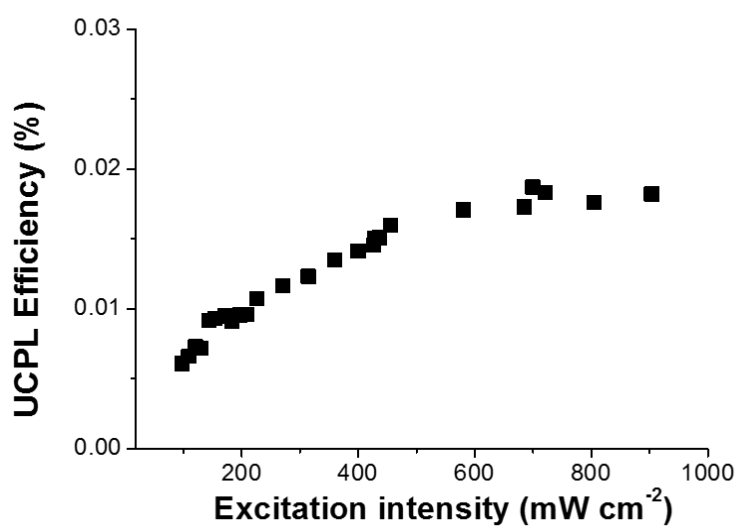

**Figure S6.** UC quantum yield $\Phi'_{UC}$ of TTM-1Cz/BPEA in toluene solution as a function of excitation intensity of the 635 laser.

**Part II. Theoretical method**

1. **Direct coupling method for doublet-triplet energy transfer integral**

For a dimer which happens doublet-triplet energy transfer

$$^2D^* + {}^1A \rightarrow {}^2D + {}^3A^* \qquad (S1)$$

The energy transfer integral of the energy transfer process:

$$V_{DA} = \langle \psi_{^2D^*} \psi_{^1A} | \hat{H} | \psi_{^2D} \psi_{^3A^*} \rangle$$
$$= \langle \psi_r | \hat{H} | \psi_p \rangle \qquad (S2)$$

$\psi_r$ and $\psi_p$ are the spin-localized wave functions before and after energy transfer, respectively. To simplify the problem, we assume that $\psi_r$ and $\psi_p$ are composed of the same set of core orbitals, and the differences are only in the four highest occupied spin-orbitals, as shown in Figure 1b. Thus, we can express $\psi_r$ and $\psi_p$ in Slater determinant:

$$\psi_r = |\psi_{core} \phi^\alpha_{D,SUMO} \phi^\alpha_{A,HOMO} \phi^\beta_{A,HOMO}|$$
$$\psi_p = |\psi_{core} \phi^\beta_{D,SOMO} \phi^\alpha_{A,HOMO} \phi^\alpha_{A,LUMO}| \qquad (S3)$$

So, the coupling can be derived as

$$V_{DA} = \left[ \phi^\alpha_{D,SUMO} \phi^\beta_{A,HOMO} \| \phi^\beta_{D,SOMO} \phi^\alpha_{A,LUMO} \right]$$
$$= \left[ \phi^\alpha_{D,SUMO} \phi^\beta_{A,HOMO} | \phi^\beta_{D,SOMO} \phi^\alpha_{A,LUMO} \right] - \left[ \phi^\alpha_{D,SUMO} \phi^\alpha_{A,LUMO} | \phi^\beta_{D,SOMO} \phi^\beta_{A,HOMO} \right]$$
$$= -\left[ \phi^\alpha_{D,SUMO} \phi^\alpha_{A,LUMO} | \phi^\beta_{D,SOMO} \phi^\beta_{A,HOMO} \right] \qquad (S4)$$

It shows the same expression as triplet-triplet energy transfer integral,[4] which is completely the exchange electronic integral.

2. **Fermi's Golden Rule**

When the regime owns weak coupling, namely the interchromophore electronic interaction is much smaller than the vibrational reorganization energy ($V_{DA} \ll E_{reorg.}$), an excited chromophore relaxes to its equilibrium geometry prior to hopping to a neighboring molecule. As a result, the energy transfer rate can be expressed by Fermi's Golden Rule:

$$k_{DA} = \frac{2\pi}{\hbar} \sum_{v_D, v_{D^*}} \sum_{v_A, v_{A^*}} P_{D^*} P_A \left| \langle \Psi_{D^* v_{D^*}} \Psi_{A v_A} | H_{DA} | \Psi_{A^* v_{A^*}} \Psi_{D v_D} \rangle \right|^2 \delta(E_{D^* v_{D^*}} + E_{A v_A} - E_{A^* v_{A^*}} - E_{D v_D})$$

(S5)

$P_{D^*}, P_A$ are distribution functions describing the population of the vibrational states. $E_{A v_A}$ is the energy includes vibrational energy and electronic energy. The delta function assures the energy conservation law. Under Condon approximation,

$$\langle \Psi_{D^*v_{D^*}} \Psi_{Av_A} | H_{DA} | \Psi_{A^*v_{A^*}} \Psi_{Dv_D} \rangle = \langle \psi_{D^*} \psi_A | H_{DA} | \psi_{A^*} \psi_D \rangle \langle \Theta_{v_{D^*}} | \Theta_{v_D} \rangle \langle \Theta_{v_A} | \Theta_{v_{A^*}} \rangle \quad \text{(S6)}$$

Recast the delta function ensuring the energy conservation for energy transfer process in two parts:

$$\delta(E_{D^*v_{D^*}} + E_{Av_A} - E_{A^*v_{A^*}} - E_{Dv_D}) = \int_{-\infty}^{+\infty} dE \, \delta(E_{D^*v_{D^*}} - E_{Dv_D} - E)\delta(E_{Av_A} - E_{A^*v_{A^*}} + E) \quad \text{(S7)}$$

Define

$$D(E) = \sum_{v_D, v_{D^*}} P_{D^*} \left| \langle \Theta_{v_{D^*}} | \Theta_{v_D} \rangle \right|^2 \delta(E_{D^*v_{D^*}} - E_{Dv_D} - E) \quad \text{(S8)}$$

$$A(E) = \sum_{v_A, v_{A^*}} P_A \left| \langle \Theta_{v_A} | \Theta_{v_{A^*}} \rangle \right|^2 \delta(E_{Av_A} - E_{A^*v_{A^*}} + E) \quad \text{(S9)}$$

which are the Franck-Condon parts of donor's emission spectrum and acceptor's absorption spectrum respectively. Therefore

$$k_{DA} = \frac{2\pi}{\hbar} |V_{DA}|^2 \int_0^{+\infty} dE \, D(E) A(E) \quad \text{(S10)}$$

For doublet-triplet energy transfer process, the emission of donor is $D_1 \rightarrow D_0$ and the absorption of acceptor is $S_0 \rightarrow T_1$.

**Part III. Theoretical results**

1. **Calculation method**

The ground states and first singlet (doublet/triplet) excited states for TTM-1Cz, BPEA and DPA were optimized at B3LYP/cc-pVDZ level within Gaussian 09 program,[5] and the corresponding frequencies were achieved at the same level. The Franck-Condon parts of TTM-1Cz emission spectrum and BPEA/DPA absorption ($S_0 \rightarrow T_1$) spectra, as well as the radiative rate of TTM-1Cz were calculated by MOMAP program.[6] The doublet-triplet energy transfer integrals were programmed and implemented in NWchem 6.0. The transfer integrals were calculated at B3LYP/6-311+G(d) level which has been proved effective for triplet-triplet energy transfer exchange integral.[4] Then, the energy transfer rates could be obtained according to Eq. (S10).

Considering the up-conversion phenomena were happened in toluene solution, we need apply molecular dynamics (MD) to simulate the real situations for achieving the equilibrium configuration. Referring the experimental parameters, one TTM-1Cz molecule and 100 BPEA/DPA molecules were randomly dissolved in toluene solvents within a $10^3$ nm$^3$ cubic box with Packmol program.[7] Then MD of solution systems was performed within NPT ensemble (constant number of atoms, pressure, and temperature) in GROMACS-4.5.7.[8] Berendsen thermostat with a time-step of 1 fs was employed to regulate the temperature at 298 K. All simulations were carried out for 3 ns to achieve a fully relaxed configuration by using General Amber Force-Field (GAFF).[9]

2. **Results and discussion**

At the optimized geometries of each state for the three molecules achieved by B3LYP/cc-pVDZ, we then apply B3LYP/cc-pVDZ and M06-2X/cc-pVDZ to calculate their excitation energies. From Table S1, it is easy to find that M06-2X/cc-pVDZ can give more reasonable vertical excitation energies compared to experiments for BPEA and DPA, while B3LYP/cc-pVDZ is more suitable for TTM-1Cz. The theoretical absorption spectrum for TTM-1Cz as shown in Figure S7 shows a good agreement with experiment. Thus, we apply M06-2X excitation energies for BPEA and DPA and B3LYP excitation energies for TTM-1Cz for investigation.

**Table S1**. The lowest theoretical vertical excitation energies for TTM-1Cz, BPEA and DPA resulted from B3LYP/cc-pVDZ and M06-2X/cc-pVDZ, compared with experiments.

|  | TTM-1Cz | | | BPEA | | | DPA | | |
| --- | --- | --- | --- | --- | --- | --- | --- | --- | --- |
|  | B3LYP | M06-2X | Exp. | B3LYP | M06-2X | Exp. | B3LYP | M06-2X | Exp. |
| $S_0 \rightarrow S_1$/eV | 1.90 | 2.76 | 2.05 | 2.43 | 2.74 | 2.70 | 3.11 | 3.46 | 3.31 |
| $S_1 \rightarrow S_0$/eV | 1.40 | 2.49 | 1.82 | 2.17 | 2.40 | 2.55 | 2.61 | 2.88 | 2.91 |

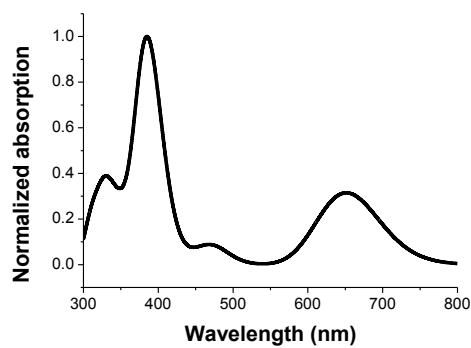

**Figure S7**. The theoretical absorption spectrum resulted from B3LYP/cc-pVDZ compared with experiments for TTM-1Cz.

Based on the excitation energy and frequencies of ground and excited states ($S_1$ and $T_1$), we adopt vibration correlation function formalism developed by Shuai *et al.* to calculate the Franck-Condon integrals of $D_1 \rightarrow D_0$ emission for TTM-Cz and and $S_0 \rightarrow T_1$ absorption for BPEA and DPA at 300K (Figure S8).[6] From Figure S8, we can see that the overlap between TTM-1Cz's emission and BPEA (DPA)'s absorption ensures the occurrence of energy transfer.

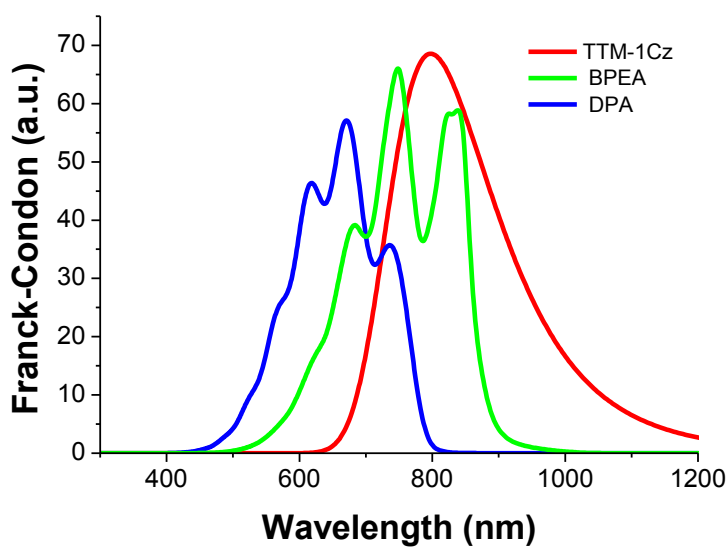

**Figure S8.** The Franck-Condon integrals of $D_1 \rightarrow D_0$ emission for TTM-1Cz and $S_0 \rightarrow T_1$ absorption for BPEA and DPA.

For simulation the real systems and calculate the energy transfer rate, we apply MD to simulate the solution systems. After reaching the equilibrium, we sampled one configuration per 0.5 ps. By counting the average centroid distance between TTM-1Cz and all BPEA (DPA) during the last 2 ns dynamic process, we find the TTM-1Cz/BPEA (TTM-1Cz/DPA) dimer with the shortest average distance. Considering the computational cost, we select one snapshot with the shortest centroid distance as example to study the energy transfer property (Figure S9). The D-T energy transfer integrals of TTM-1Cz/BPEA and TTM-1Cz/DPA dimers as shown in Figure S9 are 2.0 meV and 0.4 meV separately. Therefore, the corresponding energy transfer rates are $5.82 \times 10^{10}$ s$^{-1}$ and $5.26 \times 10^8$ s$^{-1}$ respectively. Although the shortest centroid distance between donor and acceptor does not mean the largest energy transfer rate, however, the rates are larger than the observed values ($3.18 \times 10^7$ s$^{-1}$ and $4.51 \times 10^7$ s$^{-1}$) which represent the statistical data demonstrates that D-T energy transfer can definitely happen between TTM-1Cz and BEPA/DPA. Furthermore, we also calculate the doublet exciton lifetime of TTM-1Cz with vibration correlation function method. the theoretical radiative and non-radiative decay rates of TTM-1Cz molecule without Duschinsky rotation effect are $4.83 \times 10^6$ s$^{-1}$ and $1.55 \times 10^8$ s$^{-1}$, which give rise to an exciton lifetime with 6.25 ns and agree with experiment well.

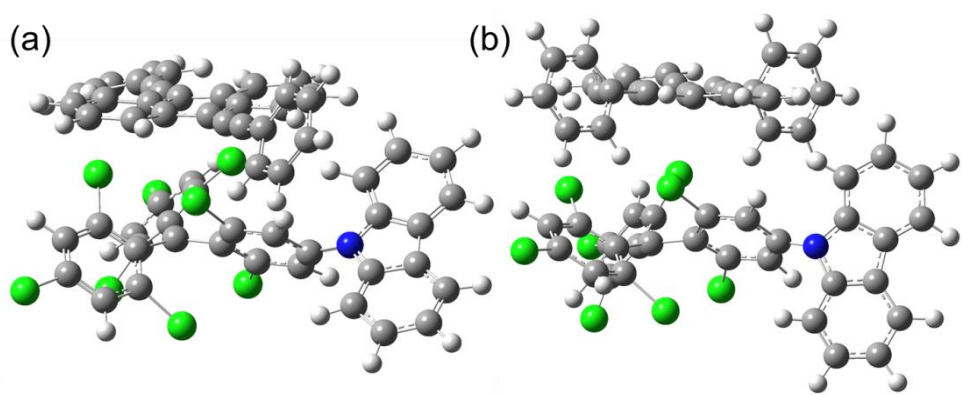

**Figure S9**. The TTM-1Cz/BPEA (a) and TTM-1Cz/DPA (b) dimers with shortest centroid distance selected from one equilibrium snapshot. The distances are respectively 5.05 Å and 5.33 Å.